\documentclass[12pt,preprint]{aastex}

\def\vec{\mathbf}
\def\ra{\rangle}
\def\la{\langle}
\def\kalias{\vec k+2k_N\vec n}
\def\rr{\rho ({\vec r})}
\def\dr{\delta ({\vec r})}

\def\kmpc{\,h{\rm Mpc}^{-1}}

\shorttitle{Power Spectrum}

\shortauthors{Cui et al.}

\begin{document}


\title{An ideal mass assignment scheme for measuring the Power Spectrum with FFTs} 

\author{Weiguang Cui\altaffilmark{1,4}, Lei Liu\altaffilmark{1,4}, 
        Xiaohu Yang\altaffilmark{1,4}, Yu Wang\altaffilmark{1,4},  
        Longlong Feng\altaffilmark{2}, Volker Springel\altaffilmark{3}} 
      \altaffiltext{1}{Shanghai Astronomical Observatory,
         the Partner Group of MPA, Nandan Road 80, Shanghai 200030, China;
         E-mail: wgcui@shao.ac.cn}
      \altaffiltext{2}{Purple Mountain Observatory, Nanjing 210008, China}
      \altaffiltext{3}{Max-Planck-Institut f\"ur Astrophysik,
            Karl-Schwarzschild-Strasse 1, 85748 Garching, Germany}
      \altaffiltext{4}{Joint Institute for Galaxy and Cosmology (JOINGC) of
        Shanghai Astronomical Observatory and University of Science and
        Technology of China}

\begin{abstract} In measuring the power  spectrum of the distribution of large
numbers of dark matter particles  in simulations, or galaxies in observations,
one   has   to  use   Fast   Fourier   Transforms   (FFT)  for   calculational
efficiency. However, because of the  required mass assignment onto grid points
in this method, the  measured power spectrum $\la |\delta^f(k)|^2\ra$ obtained
with an  FFT is not  the true  power spectrum $P(k)$  but instead one  that is
convolved with a window function $|W(\vec k)|^2$ in Fourier space. In a recent
paper, Jing  (2005) proposed an  elegant algorithm to deconvolve  the sampling
effects of  the window function  and to extract  the true power  spectrum, and
tests using  N-body simulations show that  this algorithm works  very well for
the three most commonly used mass assignment functions, i.e., the Nearest Grid
Point (NGP),  the Cloud In  Cell (CIC) and  the Triangular Shaped  Cloud (TSC)
methods. In this paper, rather  than trying to deconvolve the sampling effects
of  the  window  function, we  propose  to  select  a particular  function  in
performing  the mass  assignment that  can  minimize these  effects. An  ideal
window  function should fulfill  the following  criteria: (i)  compact top-hat
like support in  Fourier space to minimize the  sampling effects; (ii) compact
support  in real  space  to allow  a  fast and  computationally feasible  mass
assignment onto grids. We find  that the scale functions of Daubechies wavelet
transformations are good  candidates for such a purpose.  Our tests using data
from  the Millennium  Simulation show  that the  true power  spectrum  of dark
matter can be accurately measured at a level better than 2\% up to $k=0.7k_N$,
without  applying any deconvolution  processes. The  new scheme  is especially
valuable for  measurements of  higher order statistics,  e.g.~the bi-spectrum,
where it can render the mass assignment effects negligible up to
comparatively high $k$.
\end{abstract}


\section{INTRODUCTION}\label{sec_intro}

In  studies  of  the  cosmic  large-scale structure,  a  number  of  different
statistical  methods are  routinely  used to  extract  various information  of
interest (e.g.,  regarding the cosmology, the initial  perturbation, etc) that
is embedded in  the distribution of the dark matter particles  (in the case of
simulations) or the galaxies (in  observations).  The power spectrum $P(k)$ is
one of  the most  powerful and basic  statistical measures that  describes the
distribution of mass and light in  the Universe, and one of the most throughly
investigated  quantities in  modelling  the structure  formation process.  The
initial primordial power spectrum of  the mass fluctuations is usually assumed
to follow a  power law, $P_0(k)=Ak^n$.  The linearly  processed power spectrum
$P_{\rm  lin}(k)$ can  be well  predicted by  codes such  as  {\small CMBFAST}
(Seljak  \& Zaldarriaga  1996),  or approximated  by  various fitting  formula
(e.g. Bardeen  et al. 1986; Efstathiou,  Bond \& White 1992;  Eisenstein \& Hu
1998) for different matter and  energy content.  Using N-body simulations, the
non-linear power spectrum $P_{\rm NL}(k)$ has been modelled by various authors
(e.g. Peacock \&  Dodds 1996; Ma \&  Fry 2000; Smith et al.  2003). Apart from
these  theoretical  models, direct  measurement  of  the  power spectrum  from
observations plays  an extremely important  role both in cosmology  and galaxy
formation theories. Although  there are different biases relative  to the mass
power spectrum, one can roughly say  that $P(k)$ on very large scales measures
the  primordial density  fluctuations,  which is  closely  connected with  the
cosmology models  (e.g., Spergel  et al. 2007),  while $P(k)$ on  small scales
characterizes the later non-linear evolution (e.g., Peacock \& Dodds 1996).

As  an essential  statistical measure  for the  distribution of  galaxies, the
power spectrum $P(k)$ has been estimated and modeled from most of the redshift
surveys.  Recent  investigations along  this  direction  include  the CfA  and
Perseus-Pisces  redshift surveys  (Baumgart  \& Fry  1991),  the radio  galaxy
survey (Peacock  \& Nicholson 1991), the  IRAS QDOT survey  (Kaiser 1991), the
2Jy IRAS  survey (Jing \& Valdarnini  1993), the 1.2Jy IRAS  survey (Fisher et
al.  1993),  the Las  Campanas  Redshift  Survey (Lin  et  al.  1996; Yang  et
al. 2001),  the 2dF  Galaxy Redshift Survey  (Percival et al.,  2001; Tegmark,
Hamilton \& Xu  2002; S\'anchez et al. 2006) and the  Sloan Digital Sky Survey
(Tegmark et  al. 2004; Percival et  al., 2007). Among these  works, the galaxy
power  spectra are  measured either  using  the Fast  Fourier Transform  (FFT)
technique or  direct summation,  or other advanced  techniques (e.g.,  Yang et
al. 2001; Tegmark et al. 2004).

Apart  from these  observational probes,  the  power spectrum  is also  widely
measured  from  N-body  simulations  (e.g.  Davis et  al.   1985).  For  these
measurements, one  has to use FFTs since  there are too many  particles in the
simulations  to  apply  direct  summation.   Before performing  the  FFT,  one
therefore needs to assign the  particle distribution $\rho(\vec r)$ onto grids
$\rho(\vec  r_g)$  (usually  onto  $2^{3i}$   grid  cells,  where  $i$  is  an
integer). As pointed out in a  recent paper by Jing (2005), such an assignment
process is equivalent to a convolution  of the real density field with a given
assignment  window function $W(\vec  r)$, and  sampling the  convolved density
field at the $2^{3i}$ grid points. Thus the power spectrum based on the FFT of
$\rho(\vec r_g)$ is  not equal to that based on the  Fourier transform (FT) of
$\rho(\vec r)$. In order to obtain the true power spectrum to an accuracy of a
few percent,  the sampling effects  should be carefully corrected  (Jing 2005;
and references therein).

To  this  end,  Jing  (2005)  proposed an  elegant  algorithm  to  iteratively
deconvolve  the  power  spectrum  measurement  for  the  impact  of  the  mass
assignment  and  to extract  the  true  power  spectrum.  Tests  using  N-body
simulations  show that  their algorithm  works  extremely well  for the  three
commonly used mass  assignment functions, i.e., the Nearest  Grid Point (NGP),
the Cloud In Cell (CIC) and the Triangular Shaped Cloud (TSC) methods.

In this paper,  rather than trying to correct for the  influence of the window
function, we seek to minimize the  effects of the mass assignment by selecting
special  window  functions.  An  ideal  window  function  should  fulfill  the
following criteria: (i) compact top-hat like support in Fourier space to avoid
the  sampling   effects;  (ii)  compact   support  in  real  space   to  allow
computationally efficient mass  assignment onto grids. We find  that the scale
functions of  the Daubechies wavelet  transformations are good  candidates for
simultaneously  matching both requirements.  In fact,  as we  will demonstrate
they allow an accurate measurement of the power spectrum with FFTs without the
need  for a  deconvolution procedure.  This is  of great  help  especially for
accurate  measurements of higher  order spectra,  like the  bi-spectrum, where
FFTs are needed  but the de-aliasing   methods are not  available yet. We will
discuss  this  application  to  accurate  measurements and  modelling  of  the
bi-spectrum in a subsequent paper.

This paper is organized as follows.  In Section \ref{sec_form} we give a brief
description  of the  methodology for  measuring  the power  spectrum from  the
discrete distribution  of dark matter  particles.  In Section~\ref{sec_window}
we first present  the commonly used window functions in  both real and Fourier
spaces,  and  then   and  introduce  our  new  mass   assignment  scheme.   In
Section~\ref{sec_test}  we  compare  the  power  spectra  extracted  from  the
Millennium Run using different methods.   Finally, we summarize our results in
Section~\ref{sec_summ}.

\section{Measuring the power spectrum}\label{sec_form}

In this section we outline the methods used to measure the power spectrum from
the  distribution  of dark  matter  particles  (Peebles  1980). Unless  stated
otherwise we shall follow Jing (2005), and we refer readers who are interested
in a more  detailed and complete set of formulae to  this paper and references
therein.  We start from the definition of the power spectrum. Let $\rr$ be the
cosmic density  field and $\overline\rho$  the mean density. Then  the density
contrast $\dr$ can be expressed as,
\begin{equation}
\dr={\rr-\overline\rho \over \overline\rho}.
\end{equation} Based on the  cosmological principle, we assume that $\rho(\vec
r)$  in a  very  large volume  $V_\mu$  fairly represents  the overall  cosmic
density field, and that it can be taken to be periodic. The FT of $\delta(\vec
r)$ can be defined as:
\begin{equation}
\delta ({\vec k})={1\over V_\mu}\int_{V_\mu}\dr e^{i{\vec r}\cdot{\vec
    k}}{\rm d}{\vec r}\,. 
\end{equation}
And by definition, its power spectrum $P(k)$ is simply related to ${\delta
  ({\vec k})}$ as
\begin{equation}
  P(k)\equiv \la\mid{\delta ({\vec k})}\mid^2\ra \,, 
\end{equation}
where $\la\cdot\cdot\cdot\ra $ means the ensemble average.

However,  in practice,  the  cosmic density  field  is usually  traced by  the
distribution  of galaxies  or  dark  matter particles.   In  these cases,  the
density field $\rr$ is replaced  by the number density distribution of objects
$n(\vec  r) =\sum_j  \delta^D(\vec  r -\vec  r_j)$,  where $\vec  r_j$ is  the
coordinate  of  object  $j$  and  $\delta^D(\vec  r)$  is  the  Dirac-$\delta$
function.  And the FT of  the related number density contrast $\delta(\vec r)$
can be expressed as,
\begin{equation}
{\delta^d ({\vec k})}
={1\over V_\mu\overline n}\int_{V_\mu} n(\vec r) 
e^{i{\vec r}\cdot{\vec k}}{\rm d}{\vec r}-\delta^K_{{\vec k},{\vec 0}}\,,
\end{equation} where $\overline n$ is the mean number density, the superscript
$d$  represents  the {\it  discrete}  case of  $\rr$,  and  $\delta^K$ is  the
Kronecker delta. If  we divide the volume $V_\mu$  into infinitesimal elements
$\{{\rm d}V_i\}$ within which there are  either 0 or 1 objects, then the above
equation can be written as:
\begin{equation}\label{eq:deltak_d}
{\delta^d ({\vec k})}={1\over N}\sum_i n_i 
e^{i{\vec r}_i\cdot{\vec k}}-\delta^K_{{\vec k},{\vec 0}}\,,
\end{equation} where $N$  is the total number of objects  in $V_\mu$ and $n_i$
is either  0 or  1. After a  bit of algebra,  it is  seen that the  true power
spectrum can be measured via
\begin{equation}\label{eq:p_d}
P(k)\equiv \la\mid{\delta ({\vec k})}\mid^2\ra \
=\la \mid{\delta^d ({\vec k})}\mid^2\ra -{1\over N}\,.
\end{equation} Obviously, when the FT  is directly applied to the distribution
of  the galaxies  or  dark matter  particles,  one needs  to  correct for  the
discreteness (or shot noise) effect, which introduces an additional term $1/N$
to the power spectrum $\la \mid{\delta^d ({\vec k})}\mid^2\ra$.

The above method of using a direct  summation in the FT can be used to measure
the  power spectrum  from the  distribution of  galaxies, when  the  number of
objects is  not very large. However,  because of the huge  number of particles
involved in N-body  simulations, it is almost impossible to  be applied to the
dark   matter   particles  of   cosmological   density   fields.  Instead,   a
computationally attractive approach is to  use an FFT. The density contrast in
Fourier space using a FFT is,
\begin{equation}\label{eq:deltak_f}
\delta^{f} ({\vec k})={1\over N}\sum_{\vec g} n^f({\vec r}_g)
e^{i{\vec r}_g\cdot{\vec k}}-\delta^K_{{\vec k},{\vec 0}}\,,
\end{equation} 
where  the superscript  $f$  represents  the FFT.   $n^f({\vec
r}_g)$ is the convolved density value  on the $\vec g$-th grid point $\vec r_g
=\vec g H$ (where $\vec g$ is an integer vector; $H$ is the grid spacing),
\begin{equation}
n^f({\vec r}_g)=\int n({\vec r})W(\vec r-{\vec r}_g) \, {\rm d}{\vec r}\,,
\end{equation} 
where  $W(\vec r)$ is  the mass assignment function.  Note that
Eqs.~(\ref{eq:deltak_d})  and (\ref{eq:deltak_f})  are different  in  that the
summations carried  out in  the former  equation is over  the objects  and the
latter over space (the grid points). After several steps (see also Hockney and
Eastwood 1981), Jing (2005) derived the following power spectrum estimator,
\begin{eqnarray}\label{eq:p_f}
\la |\delta^{f} ({\vec k})|^2\ra &=& 
\sum_{\vec n} |W(\kalias)|^2P(\kalias) \nonumber \\
 &+& {1\over N}\sum_{\vec n}|W(\kalias)|^2\,,
\end{eqnarray} 
where $W(\vec k)$ is the FT of the window function $W(\vec r)$,
$k_N=\pi/H$  is the  Nyquist  wavenumber, and  the  summation is  over all  3D
integer vectors $\vec n$. According to equation~(\ref{eq:p_f}), one can easily
identify  the  impact   of  the  mass  assignment  onto   the  measured  power
spectrum. First,  the mass assignment  introduces the factor $W^2(\vec  k)$ to
both the  true power spectrum  and the shot  noise ($1/N$) terms.  Second, the
quantity $\la |\delta^{f} ({\vec k})|^2\ra$ is a measure for a {\it convolved}
power  spectrum  (i.e.,  the sums  over  $\vec  n$)  which suffers  from  {\it
sampling} effects. As pointed out in  Jing (2005) and will be shown in Section
\ref{sec_window},  the  sampling  effects  are significant  near  the  Nyquist
wavenumber $k_N$ and  should be carefully corrected in  an accurate measure of
the power spectrum.

\section {The role of the mass assignment function }\label{sec_window}

As shown by equation (\ref{eq:p_f}), the mass assignment window function plays
an important  role in measuring the  power spectrum using an  FFT. We separate
its impact  into two parts:  one on  the shot noise  term (second term  of the
r.h.s. of  Eq. \ref{eq:p_f})  and the  other on the  true power  spectrum term
(first term  of the  r.h.s. of  Eq. \ref{eq:p_f}). Hereafter  we refer  to the
impact  on  the  true power  spectrum  term  as  the {\it  sampling  effects}.
Usually, the impact on the shot noise term can be handled analytical according
to the  FT of the window  function.  However, because of  the convolution with
the true power spectrum, the sampling effects can not be corrected easily.

There are basically two strategies  for handling the sampling effects. One can
either  try to  correct for  them  by deconvolving  the impact  of the  window
function (which is carried  out in Jing 2005) or try to  use an optimal window
function that minimizes  the sampling effects from the  outset (the purpose of
this work).  Below we discuss a  few commonly used window functions as well as
the particular mass assignment proposed  here both in real and Fourier spaces,
and then discuss their impact on measuring the true power spectrum with an FFT
in detail.

\begin{figure*}
\includegraphics[width=1.0\textwidth]{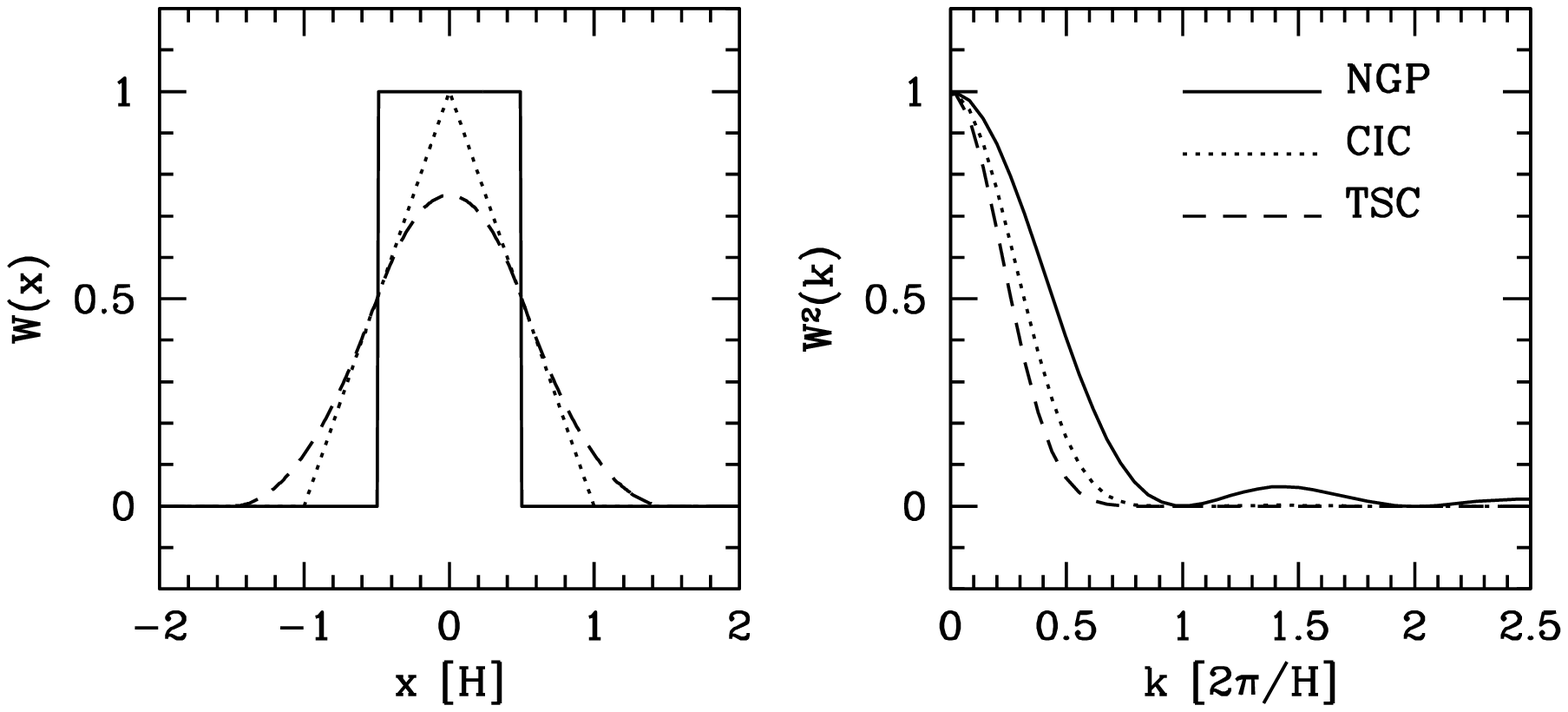}
\caption[the  mass  assignment]{Left  panel:  the  three  commonly  used  mass
assignment window functions, NGP, CIC, and TSC, as indicated. Right panel: the
square of the window functions in Fourier space.  }
\label{fig:assign_1}
\end{figure*}

\subsection{Traditional mass assignment functions}\label{sec:win1}

The  most  popular mass  assignment  functions  used  in measuring  the  power
spectrum are the  NGP, CIC and TSC  methods. Their forms in real  space can be
described by $W(\vec x)= \Pi_{i}W(x_i)$, with
\begin{equation}
W(x_i)=  \cases {1  &{$|x_i|<0.5$}\cr
0 &{else}\cr} ~~~~~~~
{\rm NGP,}
\end{equation}
\begin{equation}
W(x_i)= \cases {1-|x_i| &{$|x_i|<1$}\cr
0 &{else}\cr}~~~~~
{\rm CIC,}
\end{equation}
and
\begin{equation}
W(x_i)= \cases {0.75-x_i^2 &{$|x_i|<0.5$}\cr
(1.5-|x_i|)^2\over 2 &{$0.5<|x_i|<1.5$}\cr
0 &{else}\cr}~~~~~
{\rm TSC,}
\end{equation} 
where $x_i$ ($i=1,2,3$) is the $i$-th component of $\vec x$. In the left panel
of Fig.~\ref{fig:assign_1}, we show these window functions in real space, with
solid, dotted and dashed lines corresponding  to the NGP, CIC and TSC methods,
respectively. Their impact  on the measurement of the  power spectrum using an
FFT (Eq.  \ref{eq:p_f}) can be understood  most easily based  on their Fourier
space behavior.   According to  Hockney \& Eastwood  (1981), these  three mass
assignment window functions can be described  in Fourier space by $W(\vec k) =
\Pi_{i}W(k_i)$, with
\begin{equation}
W(k_i)=\Bigl[{\sin({\pi k_i\over 2k_N}) \over ({\pi k_i\over 2k_N})}
\Bigr]^p\,,
\end{equation} 
where $k_i$  ($i=1,2,3$) is the  $i$-th component of  $\vec k$, and  $p=1$ for
NGP,  $p=2$ for  CIC,  and $p=3$  for  TSC.  We  show in  the  right panel  of
Fig.~\ref{fig:assign_1}  (the  square  of)  the related  window  functions  in
Fourier  space. These  window  functions  peak at  $k=0$  with $W^2(k)=1$  and
decrease   sharply  with   $k\ga  0$,   especially   for  the   CIC  and   TSC
kernels. According  to Eq.~(\ref{eq:p_f}), the impact of  the window functions
can be  separated into two parts,  one on the shot  noise and one  on the true
power spectrum.

It is quite easy to model and correct the impact on the shot noise term,
\begin{equation}
D^2(\vec k)\equiv {1\over N}\sum_{\vec n}W^2(\kalias)\,.
\end{equation} 
For the  NGP, CIC, and TSC assignments,  the shot noise term  can be expressed
as,
\begin{equation}
D^2(\vec k)= {1\over N} \cases {1, &NGP, \cr
\Pi_{i}[1-{2\over 3}\sin^2({\pi k_i\over 2k_N})], &CIC, \cr
\Pi_{i}[1-\sin^2({\pi k_i\over 2k_N})+
{2\over 15}\sin^4({\pi k_i\over 2k_N})].
&TSC.\cr}
\end{equation}
In practice, for the latter two  cases, one often uses the following isotropic
approximation to model the shot noise term,
\begin{equation}
D^2(\vec k)\approx {1\over N}  \cases {[1-{2\over 3}\sin^2({\pi k\over 2k_N})],
&CIC, \cr
[1-\sin^2({\pi k\over 2k_N})+{2\over 15}\sin^4({\pi k\over 2k_N})],
&TSC,\cr}
\end{equation} 
where  $k=|\vec k|$.  As  has been  shown in  Jing (2005),  this approximation
works very  well for $k \le 0.7  k_N$, however, it can  underestimate the true
value  by about  40\% at  $k\sim k_N$.   Nevertheless, compared  to  the power
spectrum term we are  trying to measure in a CDM cosmology,  this error in the
shot noise term is usually negligible.

Now we  turn to the impact  of the window functions  on the first  term of the
r.h.s  of Eq.~(\ref{eq:p_f}), the  sampling effects.  There are  three aspects
that need to be taken into account  in measuring the true power spectrum if an
accuracy of a few {\it percent} is required.
\begin{itemize}
\item Smearing effect

  In the  summation of the  true power spectrum  over $\vec n$, only  the term
$\vec n  =0$ is what we intend  to measure. However, according  to the results
shown  in  the  right  panel  of Fig.~\ref{fig:assign_1},  the  $W^2(k)$  term
decreases sharply from $W^2(0)=1$ at $k\ga  0$, especially for the CIC and TSC
methods.  Thus the  contribution from  the {\it  related} true  power spectrum
$P(\vec k)$  is greatly suppressed. This  effect is the  so-called smearing or
smoothing effect, which  has been discussed in the  literature (e.g., Baumgart
\& Fry 1991; Jing \& Valdarnini 1993; Scoccimarro et al. 1998)

\item Anisotropy effect

  In  pratice,  one  may  use  the  average of  the  $\la  |\delta^{f}  ({\vec
k})|^2\ra$ over  different directions  for a given  $k$ to estimate  the power
spectrum $P(k)$. However,  the window function $W^2(\vec k)$  is not isotropic
for different directions for a given $k$, that is, $W^2(\vec k)$ is different,
e.g.,     for      $\vec     k=k(1/\sqrt{3},1/\sqrt{3},1/\sqrt{3})$,     $\vec
k=k(1/\sqrt{2},1/\sqrt{2},0)$,  $\vec k=k(1,0,0)$, etc.  This effect  is small
for  the NGP  method,  but quite  significant  for the  CIC  and TSC  methods,
especially at $k\sim k_N$ (eg. Baumgart \& Fry 1991, Jing 2005).

\item Aliasing effect

  The power spectrum estimator $\la |\delta^{f} ({\vec k})|^2\ra$ contains not
only  the contribution  from  $P(\vec k)$  where  $\vec n  =0$  but also  from
$P(2k_N\vec n +\vec  k)$ where $\vec n\neq 0$.  The latter contribution, which
is called the alias effect, prevents us from obtaining the true power spectrum
$P(\vec  k)$  straightforwardly. This  effect,  which  is  prominent near  the
Nyquist wavenumber  $k_N=0.5 \times (2\pi/H)$ (significant for  NGP method and
less  significant for TSC  method), has  been discussed  and handled  using an
iterative correction method in Jing (2005).

\end{itemize}

The smearing and anisotropy effects are easy to be corrected. For instance,
one can directly normalize the density contrast in Fourier space,
$\delta^f({\vec k})$, with the window function $W({\vec k})$ (e.g., Baumgart
\& Fry 1991).  Thus, the corrected density contrast $\delta^f({\vec
  k})/W({\vec k})$ obviously no longer suffer from these two effects at $k\le
k_N$, however at the price of a much enhanced aliasing effect (i.e., the
${\vec n}\ne 0$ terms in Eq.  \ref{eq:p_f}). Because of the aliasing effect
the power spectrum measured at $k=k_N$ can become a factor of 2 larger than
the true value.  Such kind of aliasing effects also exist in radio imaging
analyses based on FFTs, and various perticular mass assignment schemes have
been discussed in order to minimize their impact (e.g. Briggs et al. 1999).

Using  an  elegant iterative  correction  methods,  Jing  (2005) has  properly
corrected the  impact of the aliasing  effect, and illustrated  its success in
obtaining the true  power spectrum. On the other hand, his  method can only be
applied to the  estimation of the power spectrum.   For measurements of higher
order spectra, e.g. the bi-spectrum, there is so far no straightforward method
that can correct  the aliasing effect. In what follows,  rather than trying to
correct the above three kinds of  effects, we attemp to find a mass assignment
window function that  does not or only to very small  degree suffer from these
effects.

\begin{figure*}
\includegraphics[width=1.0\textwidth]{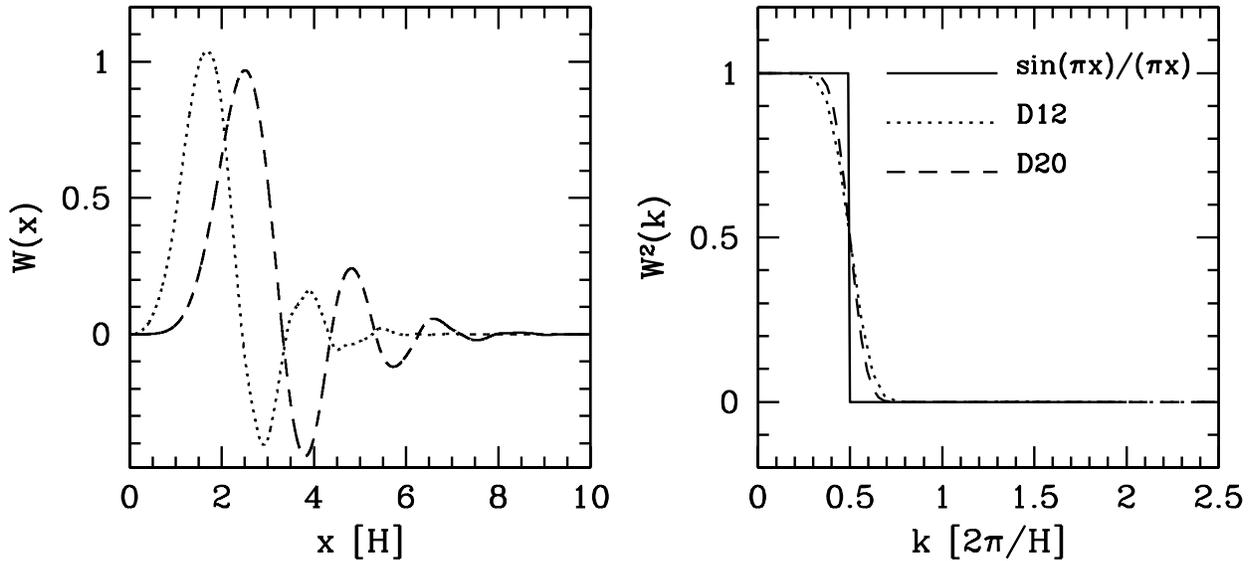}
\caption[the assignment  of Daub]{Left panel:  the scaling functions  12 (D12)
and 20  (D20) of  Daubechies, used here  as mass assignment  window functions.
Right panel: the  square of these two scaling functions  in Fourier space. For
comparison, we  also show a  top hat window  function in Fourier  space, which
corresponds to $W(x)=\sin(\pi x)/(\pi x)$ in real space. }
\label{fig:assign_2}
\end{figure*}

\subsection{Daubechies window functions}\label{sec:win2}

An  ideal window  function  that does  not  suffer from  the sampling  effects
mentioned above is  obviously a top-hat in Fourier space.  Using such a window
function, the power spectrum  measurement Eq.~(\ref{eq:p_f}) can be reduced to
Eq.~(\ref{eq:p_d}).  However  such a mass  assignment function, $W(x)=\sin(\pi
x)/(\pi x)$,  is not a compact localized  function in real space.  In the mass
assignment onto the grid, one may then have to distribute each particle's mass
to too many grid cells. In fact, if  we want to maintain an accuracy of 1\% in
the mass assignment, the mass of each particle should be distributed to $60^3$
grid cells!  Such an assignment  scheme may eliminate  most if not all  of the
computational advantage that an FFT can bring us.

Thus, a suitable  mass assignment window function should  be localized both in
real and Fourier space. A good candidate that features these properties is the
scale function  of the wavelet transformation. The  wavelet transformation has
been  previously introduced  to  astrophysical studies  and  has been  applied
successfully in the analysis of various astrophysical observations (c.f., Fang
\&  Thews 1998),  e.g., on  the distributions  of galaxies  (e.g.  Martinez et
al. 1993; Fang \&  Feng 2000; Yang et al., 2001; 2002a,b;  Feng \& Fang 2004),
on the  properties of  Ly$\alpha$ absorption lines  (e.g. Pando \&  Fang 1997;
Meiksin 2000), on  the galaxy clusters (e.g., Slezak et  al. 1994; Grebenev et
al.  1995;  Gambera et  al.  1997;  Sch\"afer et  al.  2005),  etc.  Here,  we
introduce   the   scale  function   $\phi(x)$   of   the  Daubechies   wavelet
transformation for use in power spectrum measurements, which has the following
properties (e.g., Daubechies 1992),
\begin{equation}
\int \phi(x) ~{\rm d} x \equiv 1 \,,
\end{equation}
\begin{equation}\label{eq:D_1}
\sum_{n}\phi(x+n) \equiv 1\,,
\end{equation}
and its Fourier transform, $\phi(k)$, satisfies
\begin{equation}
\int \phi^2(k) ~{\rm d} k \equiv 1 \,,
\end{equation}
\begin{equation}\label{eq:D_k1}
\sum_{n}\phi^2(k+2\pi n) \equiv 1\,.
\end{equation} 
In this paper, we use the Daubechies D12 and D20 scale functions
(Daubechies 1988, 1992) as our new mass assignment window functions,
$W(x)=\phi(x)$, which are shown in the left panel of
Fig.~\ref{fig:assign_2}.  In the right panel of
Fig.~\ref{fig:assign_2}, the squares of these two window functions in
Fourier space are shown as dotted and dashed lines, as indicated. For
comparison, we also show in the right panel the ideal case of a
top-hat Fourier window function as the solid line. The D12 and D20
window functions in Fourier space $W^2(k)$ resemble the ideal case
very well, especially in the D20 case whose deviation from the ideal
case at $k=0.35$ (i.e., $0.7k_N$) is smaller than 2\%. Note that these
particular mass assignment window functions are different from the
traditional schemes, e.g. NGP, CIC and TSC in that: (1) they are not
symmetric; (2) they are not positive definite. However these two
features will not induce any undesirable consequences in our
application. First, since the overall shifting of the window function
will not impact the amplitude of $\delta(k)$, the window function
$\phi(x)$ shown in the left panel of Fig.~\ref{fig:assign_2} can be
treated as symmetric components centered at $x\sim 1.75$ and $x\sim
2.5$, respectively, with additional fluctuations at nearby grid
cells. Second, the window function needs not necessarily be positive
definite, as we are measuring the density contrast $\delta(x)$, and
even the ideal window function $W(x)=\sin(\pi x)/(\pi x)$ is not
positive definite.

Before we  turn to a  discussion of their  impact on measuring the  true power
spectrum, let us consider the computational cost for the mass assignment using
the D12  and D20 scale functions.  According to their real  space behavior, at
much better than 0.5\% accuracy, each mass particle should be distributed onto
$6^3$ (D12) or  $8^3$ (D20) grid cells, respectively, which is  a factor of 10
or 20 times more than the TSC  method with $3^3$ grid cells. However, we argue
that  this cost is  worthwhile given  the following  positive features  of the
Daubechies assignment.

First,   according   to   Eq.~(\ref{eq:D_k1}),   the  shot   noise   term   in
Eq.~(\ref{eq:p_f}) for these mass assignment algorithms is,
\begin{equation}
D^2(k)\equiv 1/N \,.
\end{equation} 

Second, by comparing the Fourier-space  behaviors of the D12 and D20 functions
with those  of the traditional mass  assignment methods, NGP, CIC  and TSC, it
becomes  clear  that  the  three  sampling  effects  of  smearing,  alias  and
anisotropy are greatly  suppressed. Moreover, for the D20  window function, if
we only measure  the power spectrum up  to $k=0.7k_N$, we do not  need to take
into account any of those three  kinds of effects explicitly, because the true
power spectrum is recovered with better than 2\% accuracy!

Another very  important aspect is  that such a  mass assignment scheme  can be
fruitfully applied  to the  measurement of the  higher-order spectra  using an
FFT. For instance, in measuring the bi-spectrum using an FFT with the D20 mass
assignment, we do  not need to consider the sampling  effects up to $k=0.7k_N$
at all, since  here the bi-spectrum can be measured  directly with an accuracy
level better than  3\%. Note that so  far there is no other  approach known to
accurately correct for the sampling  effects in measuring the bi-spectrum with
an  FFT. We  defer  an application  of  our new  technique  and a  theoretical
modeling of the higher order spectra to a forthcoming paper.

\begin{figure*}
\includegraphics[width=1.0\textwidth]{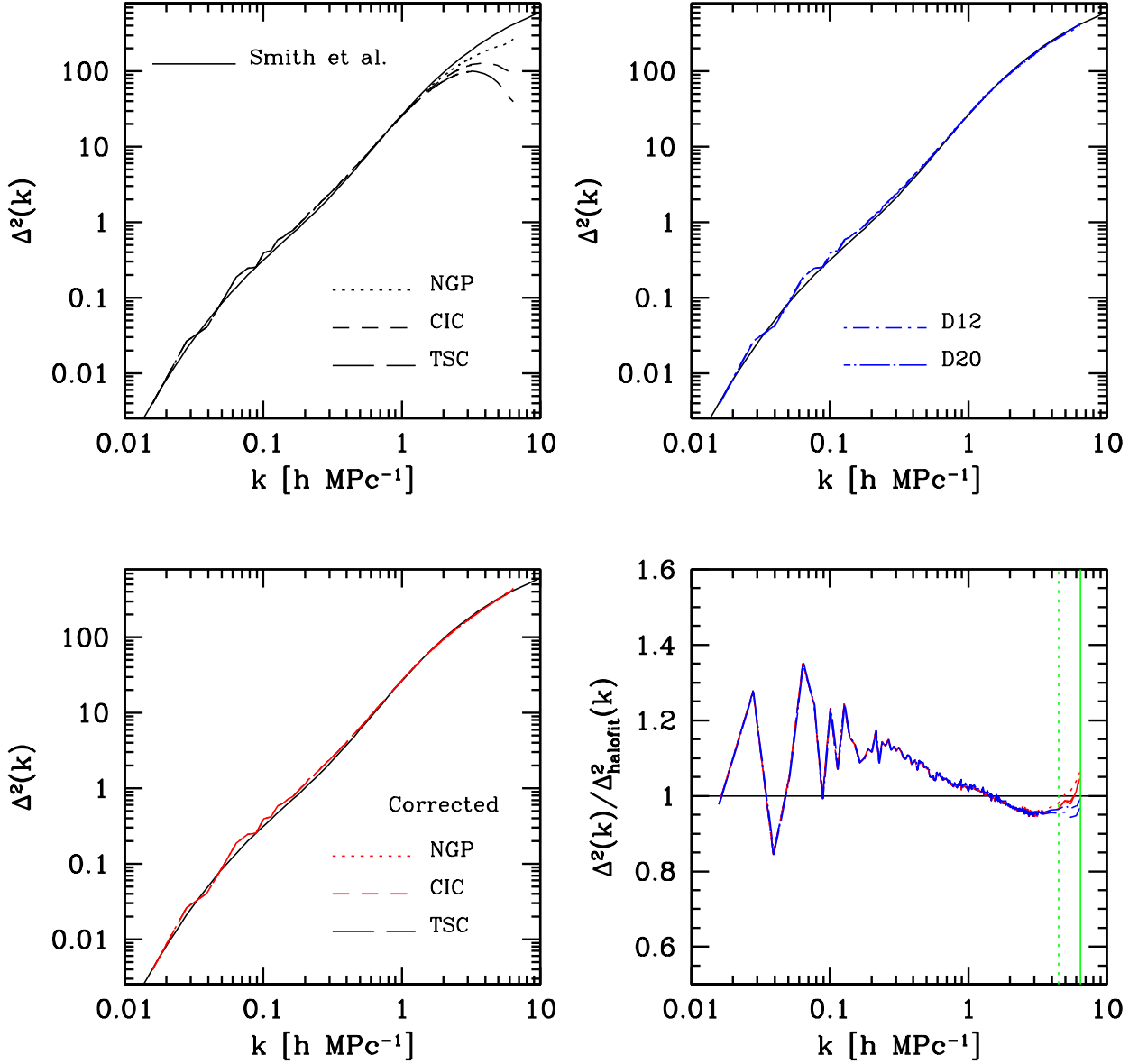}
\caption[the power spectrum]{Upper-Left panel:  the FFT power spectra measured
using the  commonly employed mass assignment  functions: NGP, CIC  and TSC, as
indicated.   In this  panel,  only the  shot  noise term  has been  corrected.
Lower-left panel: the same measurements  as shown in the upper-left panel, but
now corrected for  the sampling effects using the  iterative correcting method
proposed by  Jing (2005).  Upper-right  panel: the FFT power  spectra measured
using  the  Daubechies  scale  functions   D12  and  D20  as  mass  assignment
functions. Only the short noise term $1/N$ has been corrected.  In these three
panels, for reference  we also plot the power spectrum  prediction by Smith et
al.  (2003)  using the Millennium Run's  cosmological parameters.  Lower-right
panel: a comparison  of the ratios between the measured  power spectra and the
`halofit' prediction by  Smith et al.  (2003), for the NGP,  CIC, TSC, D12 and
D20 mass assignment  window functions.  The vertical lines  mark the locations
of $k=k_N$ and $k=0.7k_N$, respectively.  }
\label{fig:PS}
\end{figure*}

\section {Tests using N-body simulations}\label{sec_test}

Having discussed  the impact  of the mass  assignment window functions  on the
measurement of the true power spectrum using an FFT, and having introduced the
D12 and D20 scale functions,  we proceed to demonstrate their performance when
applied  to the  measurement of  the  mass power  spectrum of  a large  N-body
simulation. Here we briefly describe  the simulation, the Millennium Run, used
for this project. The Millennium Run is a very large dark matter simulation of
the  concordance   $\Lambda$CDM  cosmology  with   $2160^3\simeq  1.0078\times
10^{10}$ particles in  a periodic box of $500\,h^{-1}$Mpc  on a side (Springel
et al. 2005).  The simulation was carried out by the  Virgo Consortium using a
customized version  of the {\small GADGET2} code.  The cosmological parameters
used  in  this simulation  are  $\Omega_{\rm  m}= \Omega_{\rm  dm}+\Omega_{\rm
b}=0.25$, $\Omega_{\rm b}=0.045$,  $h=0.73$, $\Omega_\Lambda=0.75$, $n=1$, and
$\sigma_8=0.9$. For  our test  investigation, we randomly  select 10\%  of the
dark matter  particles (because  of practical limits  in computer  memory) and
measure their power spectra using  the different window functions we described
in the previous section.

To measure the power spectrum, we employ an FFT of the density distribution of
dark matter  particles assigned to a  grid with $1024^3$ cells  using the mass
assignment  algorithms   discussed  in  Section~\ref{sec_window}.    Thus  the
corresponding Nyquist wavenumber  is $k_N=1024\pi/500\,h\,{\rm Mpc}^{-1}$.  In
the  upper-left panel  of  Fig~\ref{fig:PS},  we show  the  FFT power  spectra
measured using  the traditional mass  assignment functions, NGP, CIC  and TSC,
where only the shot noise term has been subtracted.  In this figure, the power
spectrum  is presented  in terms  of $\Delta^2(k)\equiv  2\pi k^3  P(k)$.  For
comparison,  we  show  the  theoretical  prediction of  the  non-linear  power
spectrum by  Smith et al.  (2003) as  the solid line, based  on the Millennium
Run's cosmological  parameters. Obviously, because of the  sampling effects we
discussed in Section~\ref{sec:win1}, the  power spectra are quite different at
$k\ga 1\kmpc$ ($\sim 0.2 k_N$).  The power spectra measured without correcting
the   sampling  effects,   especially  for   the  TSC   method,  significantly
under-predict the  true power  spectrum.  Using the  methods proposed  by Jing
(2005),  we can  iteratively  correct  for the  sampling  effects and  extract
estimates of the true power spectrum. The corrected power spectra for the NGP,
CIC  and TSC  mass assignment  methods are  shown in  the lower-left  panel of
Fig.~\ref{fig:PS}. Comparing these power spectra among themselves and with the
`halofit' prediction of  Smith et al.  (2003), we are  convinced that the true
power spectrum  is well  recovered at  all scales $k\le  k_N$, and  is roughly
consistent with the prediction by Smith et al.

Now we  turn to use  the Daubechies  scale functions D12  and D20 as  our mass
assignment window functions. The  resulting power spectra after correcting for
the  shot   noise  term   $1/N$  are  shown   in  the  upper-right   panel  of
Fig.~\ref{fig:PS},  as  indicated. Without  any  correction  for the  sampling
effects, the measured  power spectra look very nice  and match the theoretical
predictions by Smith  et al. (2003) on  all scales up to $k\le  k_N$.  This is
very  different   from  the   results  shown  in   the  upper-left   panel  of
Fig.~\ref{fig:PS} for  the classical assignment  functions. In fact, at  a low
resolution view, there is no  visible difference between these results and the
corrected measurements shown in the lower-left panel of Fig.~\ref{fig:PS}.

Finally, we take more accurate  comparisons between the power spectra measured
with  these different  methods by  showing their  ratios with  respect  to the
`halofit' prediction of Smith et  al.  (2003).  The de-convolved power spectra
based on  the NGP, CIC and  TSC mass assignments, the  directly measured power
spectra using D12 and D20  mass assignments are ploted together for comparison
in the  bottom right panel of  Fig.~\ref{fig:PS}. Here are  a few observations
that  can be  made: (1)  the three  de-convolved power  spectra are  very well
consistent with each other at a level better than 2\% at $k\le 0.7k_N$, and at
a  level of  about 5\%  at  $k\sim 1.0k_N$;  (2) the  directly measured  power
spectra based on  the D12 and D20 (the latter  slightly better) functions have
an accuracy of better  than 2\% at $k\le 0.7k_N$ and at  a level of about 10\%
at $k\sim  1.0k_N$; (3) there is  about 20\% under-prediction  on large scales
(with $k < 1~h{\rm Mpc^{-1}}$) and  5\% over-prediction on  small scales by
Smith  et al.  (2003) for  the power  spectrum of  the  Millennium Simulation.
According  to these  findings, we  may  conclude that  both the  deconvolution
method and  the direct  measurement based on  the Daubechies  scale functions,
especially for D20, can recover the {\it true} power spectrum with better than
2\% accuracy at  $k\le 0.7 k_N$.  Moreover, as  a conservative prediction, the
bi-spectrum can be  measured at a level  better than 3\% at $k\le  0.7 k_N$ if
the D20  window function  is used in  the mass  assignment for the  FFT.  This
should be very useful for accurate studies of the bi-spectrum.

\section{SUMMARY}\label{sec_summ}

To  quantify  the  large-scale  structure  in the  distributions  of  a  large
population of dark  matter particles or galaxies, one  may measure their power
(or higher order)  spectra using a FFT. However,  the required mass assignment
onto the points of the FFT-grid can introduce sampling effects in the measured
power spectra.  Most of these effects  have been noticed and  discussed in the
literature before (e.g.,  Baumgart \& Fry 1991; Jing  \& Valdarnini 1993; Jing
2005). Among these,  Jing (2005) was the first to  use an iterative correction
method to compensate  for all of these sampling  effects, especially the alias
effect.

In  this paper,  we follow  Jing (2005)  and discuss  the impact  of  the mass
assignment  on  measuring the  power  spectrum with  an  FFT.   There are  two
components that the employed window function can impact: one is the shot noise
term and the other is the term involving the true power spectrum. With respect
to the  influence on the true  power spectrum term, there  are three different
sampling effects that need to be considered: the smearing effect, the aliasing
effect and the anisotropy effect.

Rather than trying to deconvolve for the sampling effects, we propose to use a
special  window  function: the  Daubechies  wavelet  scale  function that  can
minimize  these  sampling  effects.  In  particular, the  D12  and  D20  scale
functions considered here are compact in  real space, which allows a fast mass
assignment onto  the grid  cells, while  at the same  time their  top-hat like
shape in Fourier space leads only to very small sampling effects.

According  to the Fourier  transform $W^2(k)$  of the  D20 function,  at $k\le
0.7k_N$,  all the sampling  effects induced  by the  mass assignment  can only
affect  the measured  power spectrum  at less  than a  level of  2\%.  This is
confirmed by the tests we carried out with the Millennium Run simulation. More
importantly, as  a conservative prediction,  the new method proposed  here can
measure the bi-spectrum  of dark matter particles at  better than 3\% accuracy
for $k\le 0.7k_N$,  without the need to apply any  correction for the sampling
effects, apart from a simple substraction of the shot noise term.

\acknowledgements

We thank Yipeng Jing for helpful discussions, Olaf Wucknitz and the anonymous
referee for helpful comments that greatly improved the presentation of this
paper. This work is supported by the {\it One Hundred Talents} project,
Shanghai Pujiang Program (No.  07pj14102), 973 Program (No.  2007CB815402),
863 program (No. 2006AA01A125), the CAS Knowledge Innovation Program (Grant
No.  KJCX2-YW-T05) and grants from NSFC (Nos.  10533030, 10673023, 10373012,
10633049).

\end{document}